\documentclass{mem}
\usepackage{natbib}\usepackage{txfonts}\usepackage{balance}
\usepackage{graphicx}
\usepackage[a4paper]{hyperref}
\idline{75}{282}
\begin{document}
\def\teff{$T\rm_{eff }$}
\def\kms{$\mathrm {km s}^{-1}$}

\title{
Magnetic activity in the young star SAO~51891
}

   \subtitle{}

\author{
K. \,Biazzo\inst{1} 
\and A. \, Frasca\inst{1}
\and E. \, Marilli\inst{1}
\and E. \, Covino\inst{2}
\and J. M. \, Alcal\`a\inst{2}
\and ${\rm \ddot{O}}$. \, \c{C}akirli\inst{3}
          }

  \offprints{K. Biazzo \email{katia.biazzo@oact.inaf.it}}

\institute{
INAF - Catania Astrophysical Observatory (OACt), Catania, Italy
\and
INAF - Capodimonte Astronomical Observatory (OAC), Napoli, Italy
\and
Ege University (EUO), Astronomy Department, Bornova, Izmir, Turkey
}

\authorrunning{K. Biazzo et al.}

\titlerunning{Magnetic activity in the young star SAO~51891}
\abstract{We present preliminary results on a study based on contemporaneous photometric and spectroscopic observations 
of the young K0-1V star SAO 51891. We find that SAO\,51891, a possible member of the Local 
Association, shows emission cores in the \ion{Ca}{ii} H\&K~and fillings in 
the H$\alpha$ and \ion{Ca}{ii} Infra-Red Triplet (IRT) lines. Moreover, we detect absorption lines 
of \ion{He}{i}-D3 and \ion{Li}{i} and measure a $v\sin i$ of 19 km s$^{-1}$. A clear rotational modulation of 
both the light and the photospheric temperature, due to photospheric spots, has been detected. The net H$\alpha$ 
chromospheric emission does not show any detectable variation, while the \ion{Ca}{ii} IRT emission displays a 
fair modulation. 
}
\maketitle{}

\section{Introduction}
Stars just arrived on the Zero Age Main Sequence (ZAMS) or on their way to reach it are 
in an important evolutionary phase because they start to spin up getting free 
from their circumstellar disks which can begin to condensate giving rise to proto-planetary systems. 
At the same time, they start to loose angular momentum via magnetic braking.
SAO\,51891 is in this evolutionary stage; it is indeed a young star counterpart of an EUV source 
with an IR excess attributed to dust around the star (\citealt{Najita2005}).

%This study is a part of a project based on high-resolution spectra obtained with FOCES@CAHA (Calar Alto 
%Observatory, Spain) and SARG@TNG (Telescopio Nazionale Galileo, Spain) and contemporaneous 
%photometry performed at Catania Astrophysical Observatory (OACt, Italy) and Ege University Observatory 
%(Turkey) aimed at investigating the topology of magnetic active regions at photospheric and 
%chromospheric levels in some young single stars. 

%\section{Observations}
The spectra were acquired in August 2006 with FOCES@CAHA at a spectral resolution $R\approx40000$ in 
the wavelength range 3720-8850 \AA~with a signal-to-noise ratio higher than 200. 
The contemporaneous photometry was performed at OACt in the $BV$ Johnson bands.

%\section{Stellar parameters}
%\vspace{-.4cm}
\section{Stellar Parameters} 
Applying the ROTFIT code (\citealt{Frasca03}) to the yellow-red portion of the FOCES spectra,
we derive a spectral type of K0-1V and a $v\sin i$ of 19 km s$^{-1}$. 
From its position on the HR diagram, with the effective temperature 
$T_{\rm eff}$=5260 K derived through the line-depth ratio (LDR) method, % (cfr. Sec. \ref{sec:phot}), 
we obtain a mass of 0.8$\pm$0.2 $M_\odot$ by comparison with the evolutionary tracks of \cite{dant_mazz1997} and 
\cite{palla1999}.

%\subsection{Lithium abundance}
The lithium abundance of $\log N({\rm Li})\approx3.2$ dex deduced from the equivalent width (EW) of the Li line at 6708 \AA~
%is important because the abundance of this light element in the stellar atmosphere is related to the 
%effectiveness of the internal mixing and can be used as an `age indicator'. We find the Li abundance of SAO 51891 
is somewhat lower than the 
Pleiades upper envelope, indicating an age of $\sim100$ Myr corresponding to a Post T Tauri (PTT) or ZAMS star. 
Thus, the magnetic activity detected at photospheric (\citealt{Henry95}), chromospheric 
(\citealt{Mulliss94}) and coronal (\citealt{Voges1999}) levels should be essentially the effect 
of its young age.

%\vspace{-.4cm}
\section{Magnetic activity}
%\subsection{Diagnostics of photospheric activity}
\label{sec:phot}
\begin{figure}
\center
\includegraphics[width=3.39cm]{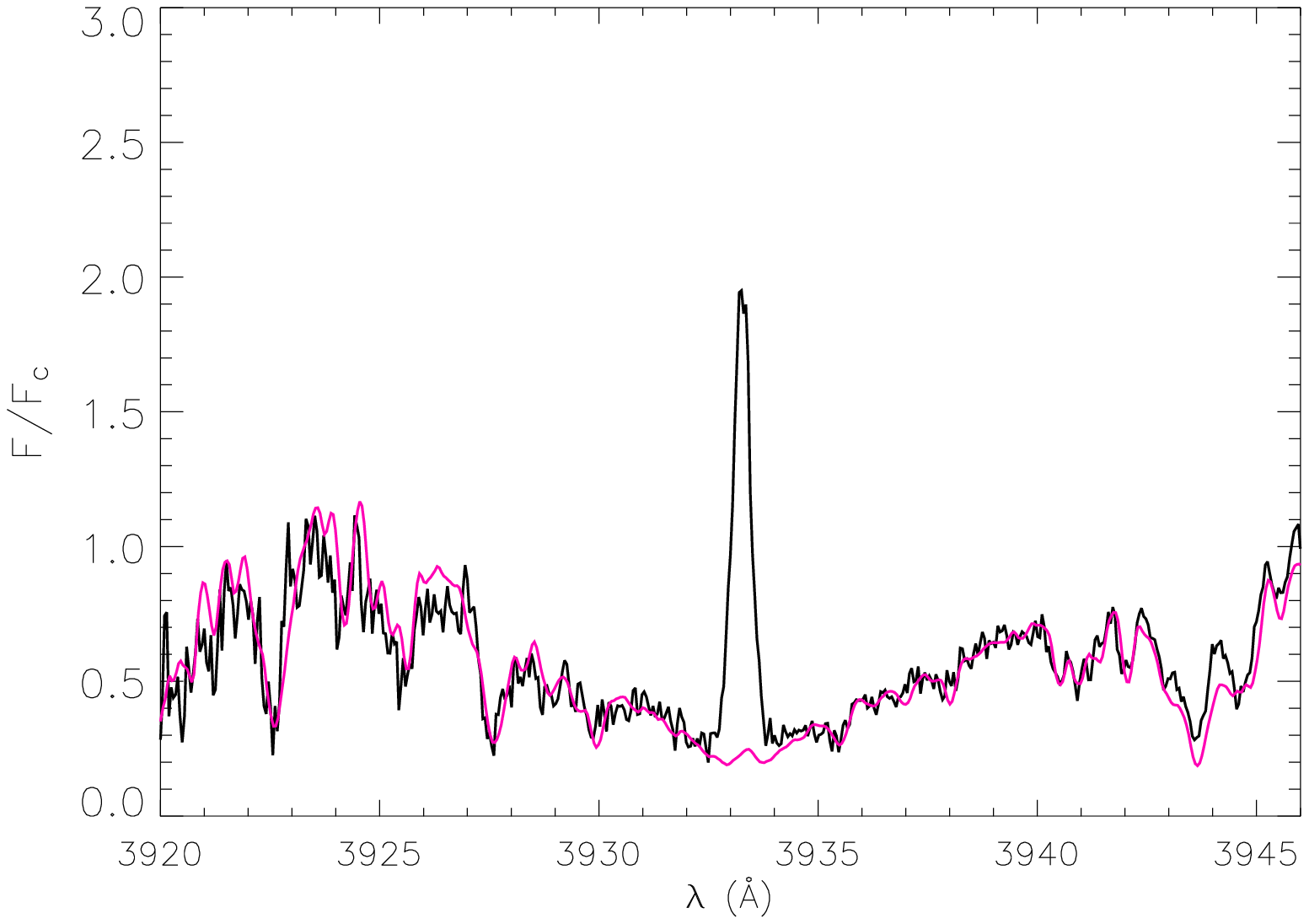}
\includegraphics[width=2.27cm,angle=90]{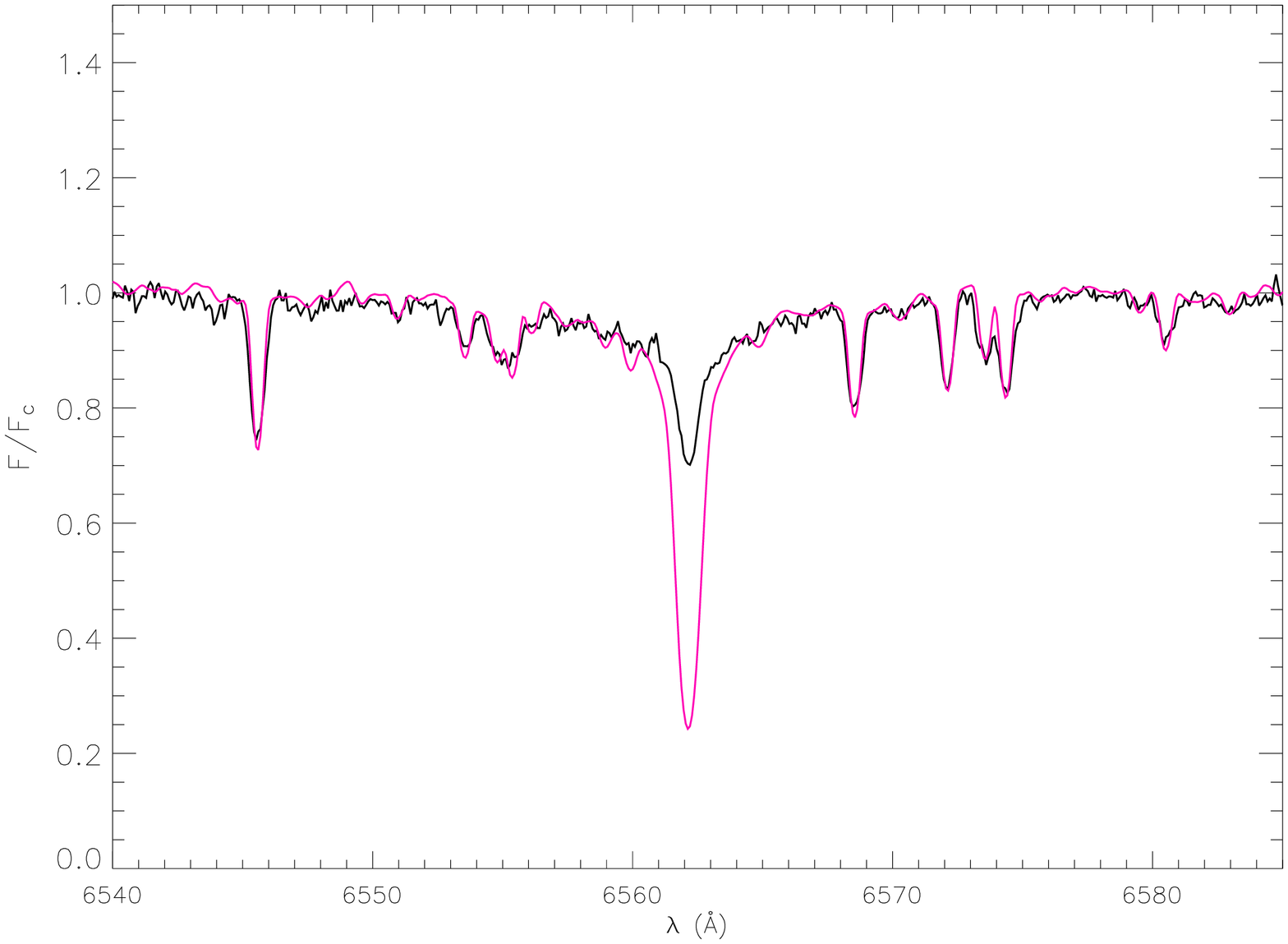}
\includegraphics[width=3.39cm]{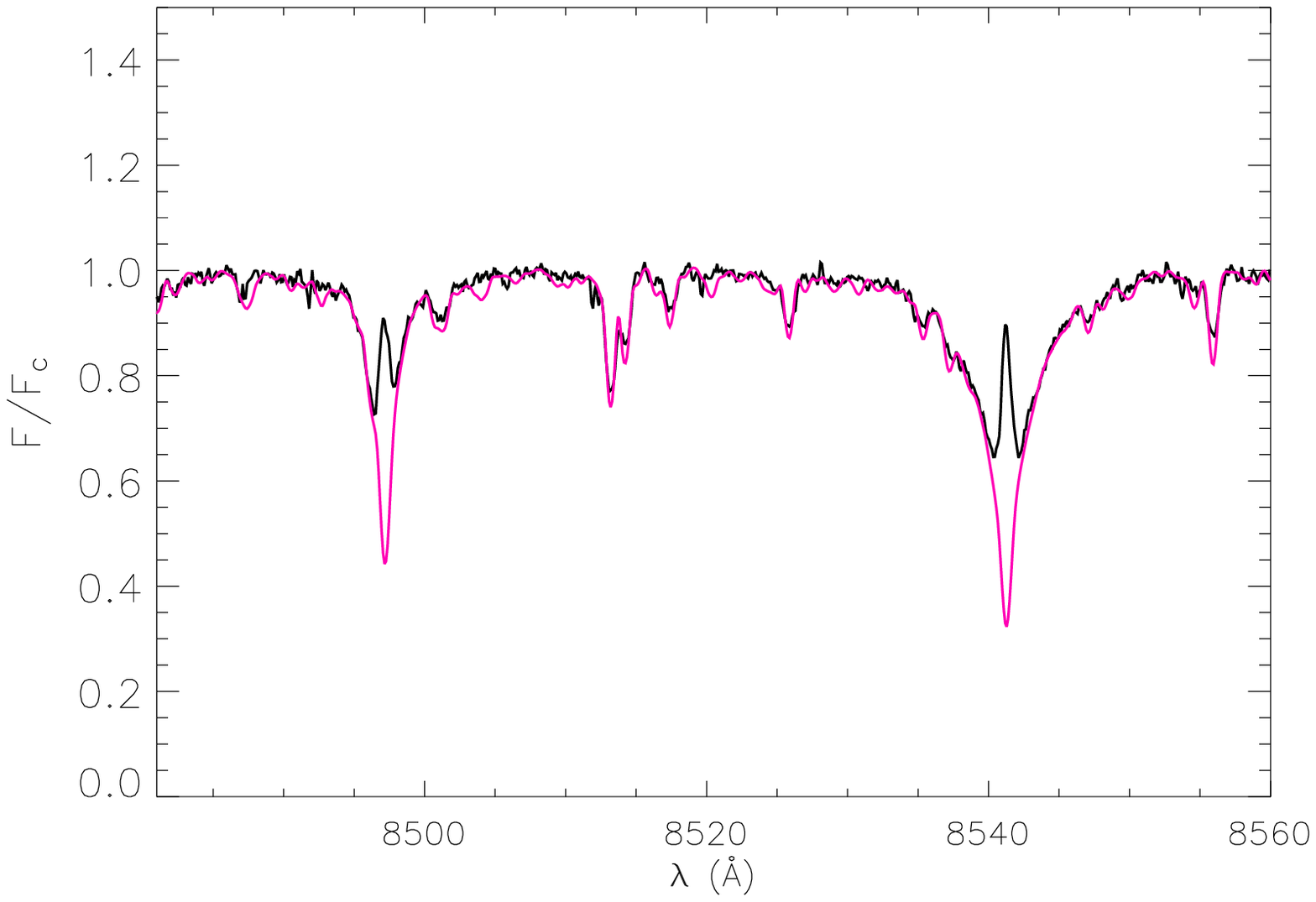}
\vspace{-.3cm}
\caption{\footnotesize 
%Example of the observed spectrum of SAO 51891 in the \ion{Ca}{ii} K, H$\alpha$, and 
%\ion{Ca}{ii} IRT (8498 and 8542 \AA~lines) regions (thick lines), together with the non-active template star 
%(thin lines).
Observed (thick lines) spectrum in three spectral regions, together with the non-active template 
spectrum (thin lines).
}
\label{fig:spectra}
\end{figure}

%As diagnostics of photospheric activity we used the $B$ and $V$  light curves and the 
The spectroscopic method based on LDRs (\citealt{Gray1991,Cata02,Biazzo07a}) allows us to detect a 
$T_{\rm eff}$ variation with an amplitude of 90\,K, which is intermediate between the value of 
$\sim\,$40 K found in stars with moderate activity (e.g., $\kappa$1 Cet; \citealt{Biazzo07b}) and 
130 K found in stars with a very high activity level (e.g., 
II Peg; \citealt{Frasca08}). Moreover, the $T_{\rm eff}$ curve is in phase with the $BV$ photometry,
confirming the hypothesis of cool spots as the primary cause of the observed variations. 

%\subsection{Diagnostics of chromospheric emission}
As diagnostics of chromospheric emission we used \ion{Ca}{ii} H\&K, \ion{He}{i}-D3, 
H$\alpha$, and \ion{Ca}{ii} IRT lines, formed at different atmospheric levels.
Using the spectral subtraction technique (\citealt{Fra94}) we obtain the chromospheric radiative losses 
in these lines. In Fig.~\ref{fig:spectra} we 
show portions of spectrum in the \ion{Ca}{ii} H\&K, H$\alpha$, and \ion{Ca}{ii} IRT spectral regions, with the 
non-active template superimposed. The H$\alpha$ and the \ion{Ca}{ii} IRT profiles are filled-in by emission, 
with the latter displaying a central reversal nearly reaching the continuum 
and suggesting a strong contribution to the total chromospheric losses (\citealt{Busa07}). The \ion{Ca}{ii} H\&K lines 
show a  strong core emission typical of cool magnetically active stars. 
Measuring the EW of residual emission profiles in the difference 
spectrum, we find that the net H$\alpha$ chromospheric emission does not show any detectable variation with 
phase, while the \ion{Ca}{ii} IRT displays a fair modulation with a possible phase shift with respect 
to the light curve.

%\vspace{-.4cm}
\section{Conclusions}
From the study of photospheric and chromospheric inhomogeneities based on spectroscopic and photometric 
monitoring of SAO 51891, we find a clear light and $T_{\rm eff}$ rotational modulation due to spots and a modulation 
of the total IRT \ion{Ca}{ii} emission due to plages. This chromospheric diagnostics seems to indicate a possible shift 
between spots and plages. Thus, as a follow/up of our previous works (\citealt{Biazzo07b,Frasca05,Frasca08}), 
we aim to develop a spot/plage model for reproducing the observed behaviors at photospheric and chromospheric 
levels and for deriving spot/plage parameters. SAO\,51891, and other 
weak-line T Tauri and PTT, and ZAMS stars already observed by us with FOCES@CAHA and SARG@TNG, 
are important to explore the correlations between global stellar parameters (e.g., 
$\log g$, $T_{\rm eff}$) and spot characteristics (e.g., filling factor and temperature) 
in stars with different evolutionary stage and activity level.

%\begin{acknowledgements}
%This work has been supported by the Italian {\em Ministero dell'Istruzione, Universit\`a e  Ricerca} (MIUR).
%\end{acknowledgements}

\bibliographystyle{aa}

\end{document}